\documentclass[%
reprint,
 amsmath,amssymb,
 aps,
]{revtex4-1}

\usepackage{float}
\usepackage{graphicx}
\usepackage{dcolumn}
\usepackage{bm}
\usepackage[flushleft]{threeparttable}
\usepackage{url}
\usepackage{amsmath}
\usepackage{amssymb}
\usepackage{braket}
\usepackage{multirow}
\usepackage{array}
\usepackage{url}
\newcolumntype{P}[1]{>{\centering\arraybackslash}p{#1}}
\usepackage{xcolor}
\usepackage{framed}
\usepackage{lipsum}
\colorlet{shadecolor}{yellow!20}
\makeatletter
\newcommand*{\rom}[1]{\expandafter\@slowromancap\romannumeral #1@}

\begin{document}

\preprint{APS/123-QED}

\title{Polarization switching mechanism in HfO$_2$ from first-principles lattice mode analysis}

\author{Yubo Qi, Sobhit Singh, and Karin M. Rabe}

\affiliation{%
Department of Physics $\&$ Astronomy, Rutgers University, \\
Piscataway, New Jersey 08854, United States
}%

\pacs{Valid PACS appear here}

\begin{abstract}

In this work, we carry out first-principles calculations and lattice mode analysis to investigate the polarization switching mechanism in HfO$_2$. Because the stability of the polar orthorhombic $Pca2_1$ phase of HfO$_2$ arises from a trilinear coupling, polarization switching requires the flipping of not only the polar $\Gamma_{15}^Z$ mode, but also at least one zone-boundary anti-polar mode. 
The coupling between the polar and anti-polar modes thus leads to substantial differences among different polarization switching paths.
Specifically, our lattice-mode-coupling analysis shows that paths in which the $X_2^-$ mode is reversed involve a large activation energy, which because the $X_2^-$ mode is nonpolar cannot be directly overcome by applying an electric field.
Our results show that the anti-polar $Pbca$ phase, whose structure is locally quite similar to that of the $Pca2_1$ phase, similarly cannot be transformed to this phase by an electric field as this would require local reversal of the $X_2^-$ mode pattern.
Moreover, for the domain wall structure most widely considered, propagation also requires the reversal of the $X_2^-$ mode, leading to a much larger activation energy compared with that for propagation of domain wall structures with a single sign for the $X_2^-$ mode.  
Finally, these first-principles results for domain wall propagation in HfO$_2$ have implications to many experimental observations, such as sluggish domain wall motion and robust ferroelectricity in thin films, and lattice mode analysis deepens our understanding of these distinctive properties of ferroelectric HfO$_2$.

\end{abstract}

\maketitle

In HfO$_2$, which is nonpolar monoclinic in bulk, the ferroelectricity observed in thin films is robust at thicknesses well below 20 nm~\cite{Muller11p112901,Muller11p114113,Olsen12p082905,Mueller12p123,Mueller12p2412,Pevsic16p4601,Cheema20p247,Muller12p4318,Polakowski15p232905,Park15p1811,Park15p192907},
making HfO$_2$-based materials of intense research interest.
The ferroelectric phase has been identified as the polar orthorhombic $Pca2_1$ phase (denoted as o-FE)~\cite{Sang15p162905,Materlik18p164101,Huan14p064111,Batra17p4139,Materlik15p134109,Batra16p172902,Xu21p1,Qi20p214108,Qi20p257603}, just one of the multiple competing polymorphs of HfO$_2$~\cite{Huan14p064111,Antunes21p082903,Materlik15p134109,Raeliarijaona21}.
Bulk HfO$_2$ adopts a cubic fluorite structure with space group $Fm\overline{3}m$ at high temperatures, and transforms to a tetragonal (denoted as t) phase~\cite{Wang92p5397}, 
a distortion of the cubic phase generated by a $X_2^-$ lattice mode, as the temperature decreases.
Recent studies show that the t phase can be induced to transform to the o-FE phase through doping, growth in thin film form, quenching, or mechanical confinement during cooling~\cite{Boscke11p102903,Boscke11p112904,Sang15p162905,Xu21p1}.
Besides the $X_2^-$ mode, the o-FE phase is stabilized by trilinear coupling of a polar $\Gamma_{15}^Z$ to additional zone-boundary anti-polar lattice distortions~\cite{Reyes14p140103,Delodovici21p064405,Raeliarijaona21}.
Further, applying hydrostatic pressure can drive a transition to the anti-polar orthorhombic $Pbca$ phase (denoted as o-AP)~\cite{Lowther99p14485,Ohtaka90p193,Jayaraman93p9205,Ohtaka01p1369}.
The structure of the o-AP phase is related to that of the o-FE phase by reversal of the polarization direction of alternate polar layers (Fig. S1).

Investigating the intrinsic nature of polarization switching in pure HfO$_2$ is experimentally challenging, since the ferroelectric phase is generally stabilized through doping in thin films, leading to local distortions and polycrystalline samples with a mixture of competing phases. 
Polarization switching in HfO$_2$ has thus attracted intensive theoretical investigation~\cite{Choe21p0,Lee20p1343,Ding20p556,Chen20p085304,Clima14p092906,Huan14p064111,Raeliarijaona21}. In particular, for the lowest-energy domain wall structure, Lee {\em{et al.}} showed that the domain-wall width is essentially zero, so polarization switching of a single layer and of the entire system requires have the same (large) energy barrier per f.u.~\cite{Lee20p1343}. For a different domain wall structure, Choe {\em{et al.}} showed that propagation would have a much lower activation energy~\cite{Choe21p0}. 
Indeed, due to the complexity of the domain wall structures and switching mechanisms, arising from the complexity of the ferroelectric phase and proposed intermediate structures, many polarization switching paths are possible candidates. The key issue is how the polarization can flip in the presence of nonpolar distortions, presenting a challenge to be addressed by a systematic approach.


In this work, we integrate lattice mode analysis with density functional theory (DFT) calculations to investigate the polarization switching mechanism in HfO$_2$ from first principles. 
The multiple modes in the o-FE phase mean the each polarization state has multiple variants, with multiple possible switching paths connecting up and down.
Our results show that polarization switching in HfO$_2$ requires the flipping of not only the polar $\Gamma_{15}^Z$ mode, but also at least one zone-boundary anti-polar mode.
Our mode coupling analysis demonstrates that this $X_2^-$ mode will not switch under an electric field due to the specific pattern it couples with the $\Gamma_{15}^Z$ mode. 
Further, we propose that the o-AP phase cannot be transformed to the polar orthorhombic phase by an electric field, because this anti-polar phase is composed of up- and down-polarized cells with opposite signs of the $X_2^-$ mode amplitudes.
We construct domain walls for the various pairs of variants and investigate the domain wall propagation energetics. 
We find that the domain wall motion in the structures with a single sign of the $X_2^-$ mode requires a much lower activation energy than for the motion of the lowest energy domain wall, which involves a change of sign of the $X_2^-$ mode.
This domain wall activation energy is still much larger than that in conventional ferroelectrics, consistent with the observation of sluggish domain wall motions~\cite{Lee20p1343,Buragohain18p222901,Mulaosmanovic17p3792,Wieder57p367,Nagarajan99p595,Lee20p1343,Son08p064101,Pantel10p084111}.
Our results demonstrate that the distinct polarization switching behavior in HfO$_2$ result from the intrinsic nature of multiple lattice modes in the ferroelectric phase, and their coupling to each other and to electric fields.

We start with an up-polarized state. Because of the nonpolar distortions in the o-FE structure, there are multiple down-polarized variants. These variants can be specified by the signs of symmetry-adapted lattice modes and are related by symmetry operations of the high-symmetry cubic structure. More specifically, as demonstrated in previous works~\cite{Reyes14p140103,Qi20p214108,Delodovici21p064405,Raeliarijaona21}, the structure of the o-FE phase can be characterized by a zone-center $\Gamma_{15}^Z$ mode, which accounts for the polarization along the $z$-direction, and four zone-boundary anti-polar $X_2^-$, $X_5^Y$, $Y_5^Z$, and $Z_5^X$ modes (see SM section~\rom{3} A).
The $X_2^-$ mode, which is the only non-zero mode in the t phase, describes the displacements of oxygen chains along the $x$-direction, alternating in a 2D checkerboard pattern. 
Applying a mirror reflection operation with respect to the $x$ or $y$ direction can change the signs of one or several anti-polar modes without changing the direction of polarization, and thus leads to a different variant. As shown in Fig. S5 and Table S3, there are four distinct variants with the same polarization (denoted as type \rom{1}$\sim$\rom{4}), which are generated by applying mirror reflections in $x$, in $y$, and in both $x$ and $y$ to the up-polarized state designated type \rom{1}.

The structural transformation paths from a specific up-polarized state to different variants of the down-polarized state are expected to be different, involving a change in sign of one or more anti-polar modes in addition to the change in sign of the $\Gamma_{15}^Z$ mode. 
To understand how the anti-polar modes evolve as the $\Gamma_{15}^Z$ mode is reversed, we begin with the o-FE structure, artificially decrease the amplitude of the $\Gamma_{15}^Z$ mode (denoted as $Q\left(\Gamma_{15}^Z\right)$) gradually, and optimize the structure with $Q\left(\Gamma_{15}^Z\right)$ fixed at each step. (Computational details can be found in Supplementary Materials (SM)~\cite{SM} section \rom{2}). 
The changes of the amplitudes of the anti-polar modes with respect to $Q\left(\Gamma_{15}^Z\right)$ are shown in Fig. S6.
In the o-FE phase, the amplitudes of the three anti-polar modes $X_5^Y$, $Y_5^Z$, and $Z_5^X$ modes are approximately the same, indicating that the non-zero $X_2^-$ and $\Gamma_{15}^Z$ modes induce little anisotropy in the lattice-mode couplings. 
We observe a first-order phase transition at $Q\left(\Gamma_{15}^Z\right)=0.108$~\AA, below which the structure transforms to the structure with the $Aba2$ space group, which can be viewed as a polarized t phase with zero $Q\left(Y_5^Z\right)$ and $Q\left(Z_5^X\right)$.
This phase transition is associated with jumps in the amplitudes of all the anti-polar modes. 
%

The behavior observed in Fig. S6 can be understood by considering the lowest-order mode coupling terms, using the cubic structure as a reference.
The t phase is generated by the $X_2^-$ mode, indicating that this mode is unstable by itself and can be characterized by an energetic term as $a_1Q\left(X_{2}^-\right)^2$, where $a_1$ is a negative coefficient (see SM section~\rom{3} C).
The $X_2^-$ mode couples to the $\Gamma_{15}^Z$ mode, leading to a biquadratic term $a_2Q\left(\Gamma_{15}^Z\right)^2Q\left(X_{2}^-\right)^2$.
This accounts for the nonzero value of $X_2^-$ when $Q\left(\Gamma_{15}^Z\right)$ is zero and its quadratic increase as $Q\left(\Gamma_{15}^Z\right)$ is increased.
In the t phase ($Q\left(X_2^-\right)$ nonzero), the polar $\Gamma_{15}^Z$ mode couples bilinearly with $X_5^Y$ through the trilinear term $a_3Q\left(X_{2}^-\right)Q\left(\Gamma_{15}^Z\right)Q\left(X_{5}^Y\right)$, where $a_3$ is a negative coefficient. Any nonzero $Q\left(\Gamma_{15}^Z\right)$ results in a nonzero $Q\left(X_{5}^Y\right)$. These two terms couple approximately linearly and become zero together, as observed in Fig. S6.
Moreover, the $\Gamma_{15}^Z$ mode has a trilinear coupling with the $Y_5^Z$ and $Z_5^X$ modes through the term $a_4Q\left(X_{2}^-\right)Q\left(\Gamma_{15}^Z\right)Q\left(Y_{5}^Z\right)Q\left(Z_{5}^X\right)$. This mixing makes $Q\left(Y_5^Z\right)$ and $Q\left(Z_5^X\right)$ both zero in the $Aba2$ phase ($Q\left(\Gamma_{15}^Z\right)<0.108$~\AA).

An electric field has the same symmetry as the polar $\Gamma_{15}^Z$ mode, so it directly couples bilinearly to this mode.
For the same reason, the coupling terms that include a given anti-polar mode and the electric field have the same form as the coupling terms for the polar $\Gamma_{15}^Z$ mode~\cite{Chen19p247701,Varignon16p057602,Yang12p057602,Juraschek17p054101,Mankowsky17p197601,Iniguez08p117201}.
Under reversal of an electric field, the sign of $Q\left(X_{2}^-\right)$ stays the same; $Q\left(X_5^Y\right)$ varies linearly with the electric field as the field goes through zero; $Q\left(Y_{5}^Z\right)$ and $Q\left(Z_{5}^X\right)$ become zero below a critical field strength and when the reversed field reaches the critical value, they become nonzero with the sign of one of them reversed. 

The nature of lattice mode coupling is thus crucial to the electric-field-driven polarization switching in HfO$_2$. Here, we divide the polarization switching paths into two categories based on whether the $Q\left(X_{2}^-\right)$ of the down-polarized state has the same, or opposite, sign as the $Q\left(X_{2}^-\right)$ in the up-polarized state. In first-category switching, the down-polarized states are type~\rom{1} or type~\rom{2}, whose $Q\left(X_{2}^-\right)$ has the same sign as the $Q\left(X_{2}^-\right)$ in the up-polarized state, and in the second-category switching, the final state should be type~\rom{3} or type~\rom{4}.

 \begin{figure*}
 \centering
\includegraphics[width=16.0cm]{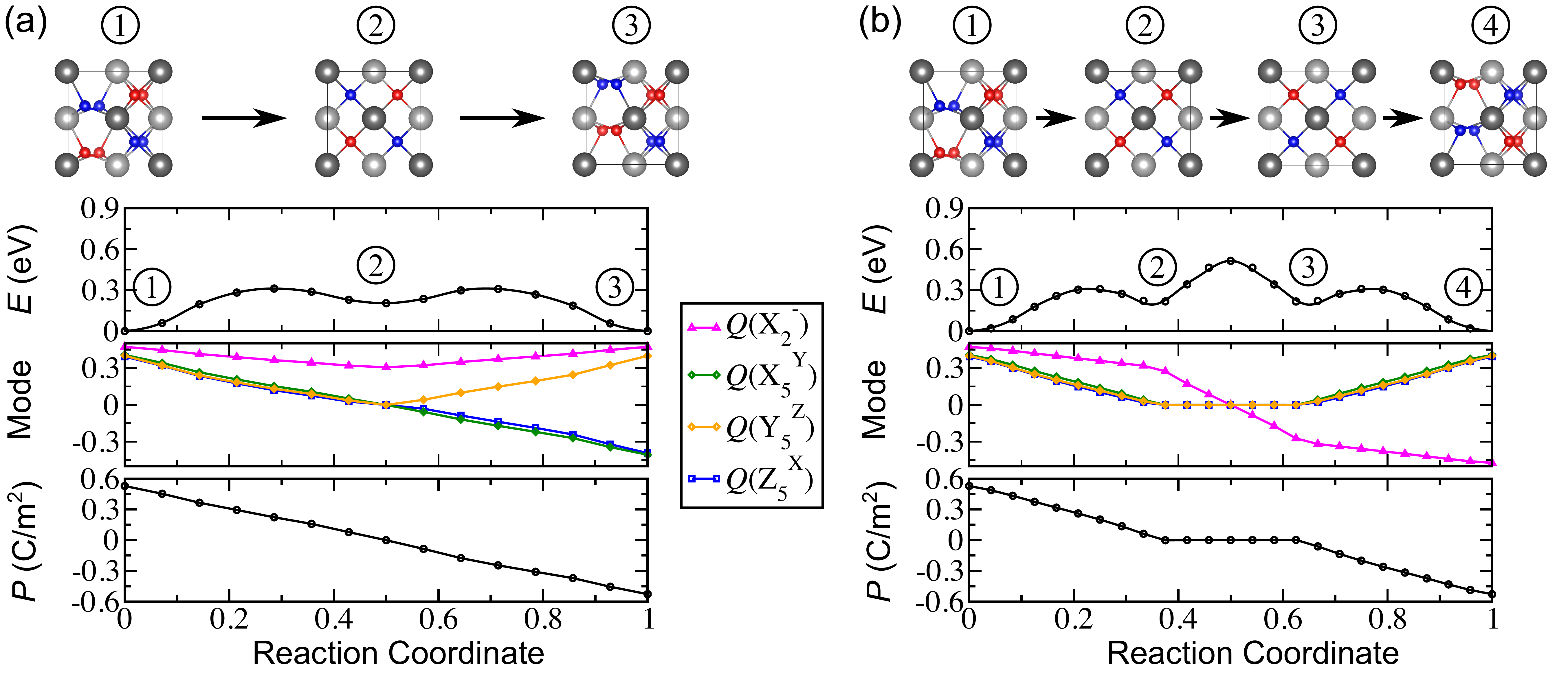}
\caption{Energy (of the unit cell composed of 4 formula units) profiles and changes of lattice-mode amplitudes during the structural changes toward the type~\rom{1} and type~\rom{3} down-polarized states. The plots for the type~\rom{2} and type~\rom{4} cases are quite similar and shown in Fig. S7. To highlight the checkerboard pattern of the oxygen $x$-direction displacements in the $X_2^-$ mode, we color the inward-displaced and outward-displaced oxygen atoms with red and blue colors respectively.}
\label{f3}
\end{figure*}

To investigate the minimum energy path and corresponding structural transformation during a uniform switching in bulk HfO$_2$, we carry out first-principles calculations with the nudged elastic band (NEB) method~\cite{Henkelman00p9901}.
In Fig.~\ref{f3} (a), we plot the minimum energy path for switching to the type~\rom{1} down-polarized state as an example of first-category polarization switching. The changes of lattice mode magnitudes, which characterize the structural transformation, are also shown.
The minimum energy path found by the NEB calculation has the t phase as an intermediate state.
Along the path, $Q\left(X_2^-\right)$ does not change much.
For the mode amplitudes ($Q\left(\Gamma_{15}^Z\right)$, $Q\left(X_{5}^Y\right)$, and $Q\left(Y_{5}^Z\right)$) whose signs are different in the up- and down-polarized states,
their values change approximately linearly through the path, which are interpolations based on the initial and final values.
On the other hand, for the mode (the $Z_{5}^X$ mode) whose amplitude does not change sign, the mode amplitude $Q\left(Z_{5}^X\right)$ decreases to zero first and then increases back to the original value, forming a `v' shape, since the t phase is an intermediate state.

In Fig.~\ref{f3} (a), we plot the minimum energy path and changes of mode amplitudes for switching to the type~\rom{3} down-polarized state as an example of second-category polarization switching.
The minimum energy path is noticeably different from that of a first-category polarization switching by having two intermediate states, which are the t phases with opposite signs of $Q\left(X_2^-\right)$.
In the middle stage, which is between the two intermediate states, $Q\left(X_2^-\right)$ decreases linearly, reverses its sign, and the amplitudes of all other modes remain the same.
Switching the sign of the $X_2^-$ mode requires extra energy, leading to a 0.514 eV energy barrier, which is 0.206 eV higher than the energy barrier during the first-category polarization switching. 

Under an electric field, $Q\left(\Gamma_{15}^Z\right)$ should change approximately linearly following the path shown in Fig.~\ref{f3} (a). The middle stage in Fig.~\ref{f3} (b), in which only $Q\left(X_2^-\right)$ changes but $Q\left(\Gamma_{15}^Z\right)$ remains zero, cannot be driven by an electric field. Therefore, we conclude that under an electric field, only first-category switching can occur, and the up-polarized state can only transform to the type~\rom{1} or type~\rom{2} down-polarized state. 

\begin{figure*}
\centering
\includegraphics[width=18.0cm]{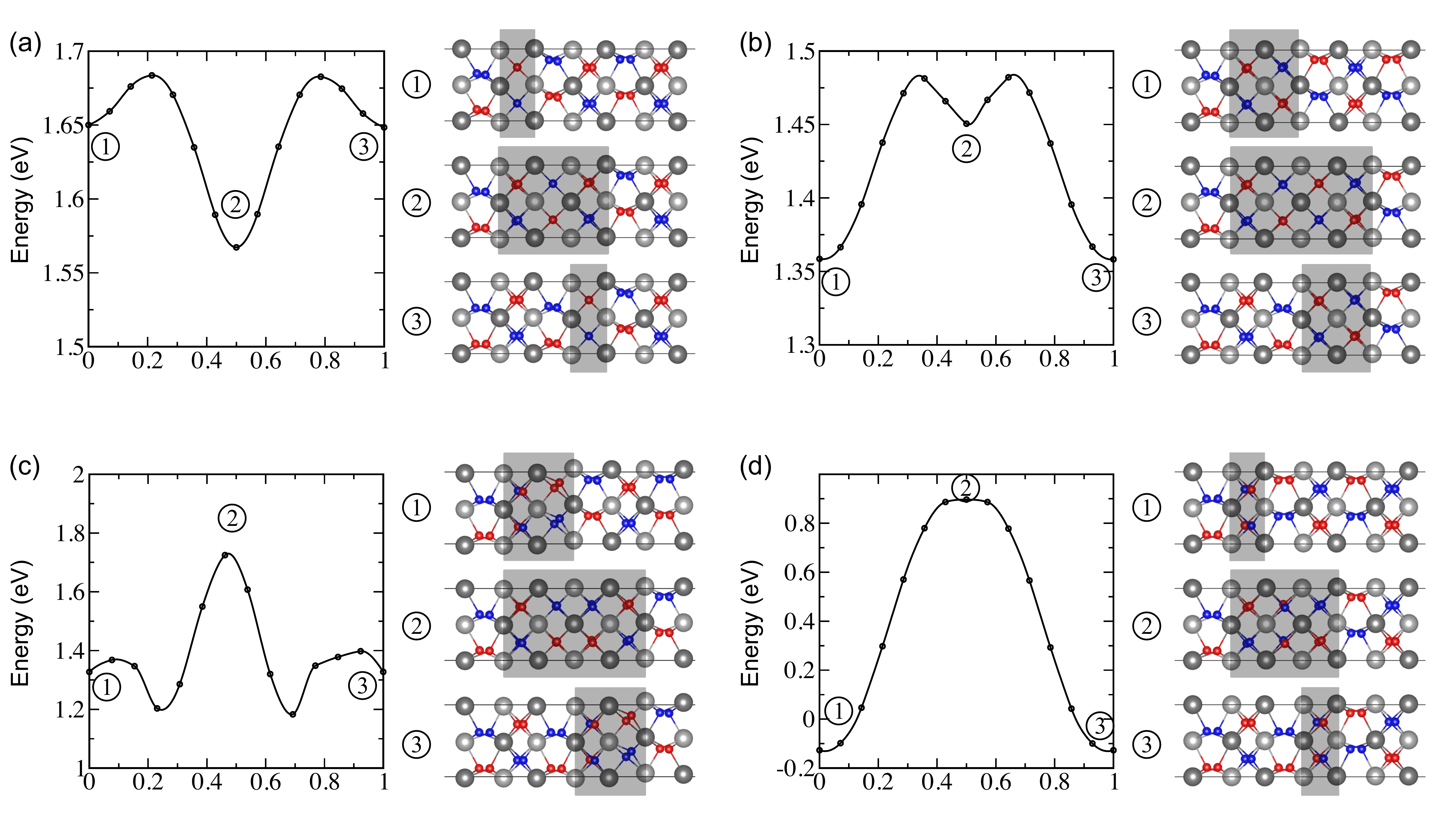}
\caption{Energy (of the entire supercell composed of 8 unit cells, 32 formula units) profiles during the domain-wall propagations. 
The shaded areas indicate the distorted parts around domain walls. 
The energy of the uniformly polarized structure is take as zero. 
Therefore, the energy of structure 1 is twice of the domain wall energy.
}
\label{f4}
\end{figure*}

In most ferroelectrics, polarization does not switch uniformly under an electric field.  
Instead, under a reverse electric field, a nucleus with opposite polarization emerges and grows, with the domain wall propagation driven by the electric field~\cite{Merz54p690,Stadler63p3255,Tybell02p097601,Ahn04p488,Li04p1174,Gruverman05p082902,Shin07p881,Liu16p360}. This is because the activation energy of domain wall propagation is usually much less than that of uniform switching. In the following part, we will investigate the domain wall propagation in HfO$_2$, showing its relationship and difference with the uniform switching.

We constructed domain structures composed of four up-polarized unit cells and four down-polarized unit cells. 
We considered four different types of domain walls (as shown in Fig.~\ref{f4}, structure 1 in the right panel of each subfigure), separating the up-polarized structure and the four different down-polarized variants. 
In type~\rom{1} and type~\rom{2}, the sign of the $X_2^-$ mode amplitude does not change across the domain wall (first category as discussed above), while in type~\rom{3} and type~\rom{4} the sign reverses (second category).
In the optimized domain structures, only the two unit cells at the domain wall are distorted, and the unit cells which are nearest neighbors to the domain wall have approximately the same structure as the uniformly polarized bulk. 
These ultra-narrow domain walls can be understood as the result of weak inter-cell lattice mode interaction, which is also manifested by the flat phonon bands of HfO$_2$~\cite{Lee20p1343} and can explain why ferroelectricity is robust even in ultra-thin films. That is, 
surface termination or bonding with electrodes can only influence the structure of the surface/interface layer, and the second layer recovers the structure and polarization of the bulk.

The energy of the optimized domain structure relative to the  uniformly up- (or down-) polarized supercell is twice the domain wall energy as there are two domain walls per supercell.
The type~\rom{1}, type~\rom{2} and type~\rom{3} domain structures have positive domain wall energies, ranging from 0.68 to 0.78 eV/cell (see SM section~\rom{6} for more discussion), while the type~\rom{4} domain structure is -0.063 eV/cell.  Next, we expand the up-polarized domain by one unit cell and contract the down-polarized domain by one unit cell (structure 3 in the right panel of each subfigure in Fig.~\ref{f4}).
The supercell energies of structure 3 for all four types of domain walls are very close to the energies of the corresponding structure 1, consistent with the observation that the inter-cell lattice mode interaction in HfO$_2$ is weak.

The negative domain wall energy for type~\rom{4} suggests that a supercell with alternating up- and down-polarized unit cells corresponds to a phase with a lower energy than the o-FE phase.
Our structural analysis shows that this anti-polar structure is exactly the Pbca o-AP phase observed at high pressures, as shown in Fig. S1. 
We note that the propagation of type~\rom{4} domain structure is widely accepted in previous theoretical work~\cite{Lee19p38929,Lee20p1343,Ding20p556} as the mechanism for polarization switching. 
In the following, we discuss alternative domain wall structures with much more physically reasonable values for the energy barrier to propagation.

To investigate the propagation of the various domain walls, we use the NEB method to calculate the minimum energy paths connecting structures 1 and 3. 
The energy profiles including initial, intermediate/transition, and final states during the domain wall propagation are shown in Fig.~\ref{f4}. %

The propagation of the domain wall depends on whether it is in the first category ($Q\left(X_{2}^-\right)$ of the down-polarized state has the same sign as the $Q\left(X_{2}^-\right)$ in the up-polarized state) or the second category (with opposite signs of $Q\left(X_{2}^-\right)$).
Our calculations show that the activation energy in the first category domain wall propagation is about 0.12 eV/cell (Fig.~\ref{f4} (a) and (b)), which is one third of the activation energy of uniform switching.
As for the case of uniform switching, there is an intermediate structure in which the unit cell at the domain wall adopts the t phase (structures 2 in Fig.~\ref{f4} (a) and (b)).
In fact, this intermediate structure in type~\rom{1} is slightly lower in energy, making it the favored domain wall in this case (see Fig. S9).

The situation is quite different in the second-category domain wall propagation. In the middle of the structural transformation path, there is a high-energy transition state, rather than a local-minimum intermediate state (Fig.~\ref{f4} (c) and (d)).
In the transition state, the two unit cells at the domain wall adopt a high-energy t-like structure which is distinct from the metastable t phase (structures 2 in Fig.~\ref{f4} (c) and (d)).
The $x$-direction displacements of oxygen atoms in the domain wall correspond to an atomically sharp change in sign of the $X_2^-$ mode pattern, matching to the different signs in the two domains.
The activation energies of the type~\rom{3} and type~\rom{4} domain wall propagations are 0.57 eV/cell and 1.02 eV/cell respectively, which are much higher than those during a first-category switching (0.12 eV/cell). 
Moreover, the structural transformation is associated with a change of the $X_2^-$ mode (indicating by the change of the arrangement of oxygen atoms colored with red and blue), which cannot be driven by an electric field as it does not involve a net change in local dipole moment.
All these results demonstrate that second-category domain wall propagation are unlikely to occur.

We conclude that switching occurs by motion of first category domain walls.
It is worth mentioning that the 0.12 eV/cell activation energy is still five times of that in PbTiO$_3$~\cite{Lee20p1343}, which explains why HfO$_2$ has a sluggish domain wall motion compared with that in conventional perovskite ferroelectrics~\cite{Buragohain18p222901,Mulaosmanovic17p3792,Wieder57p367,Nagarajan99p595,Lee20p1343,Son08p064101,Pantel10p084111}.

Another implication of this work is that the o-AP phase cannot be polarized to the o-FE phase by applying an electric field. This is because the down-polarized unit cell in the o-AP phase is the type~\rom{4} one, whose $Q\left(X_2^-\right)$ has an opposite sign compared with the up-polarized structure. Reversing $Q\left(X_2^-\right)$ not only requires a higher activation energy (as shown in Fig.~\ref{f3} (b)), but also cannot be driven by an electric field (see SM section~\rom{7}). 

In summary, we carry out first-principles calculations and lattice mode analysis to investigate the polarization switching mechanisms in HfO$_2$. We demonstrate that the polarized state has different variants due to multiple lattice modes generating the o-FE phase. 
Our mode-coupling analysis demonstrates that the sign of the anti-polar $X_2^-$ mode is the key factor, since flipping this mode not only requires a larger energy, but also cannot be directly accomplished by applying an electric field.
These results imply that the high-pressure Pbca anti-polar phase can not be transformed to the polar phase by applying an electric field, due to the alternating sign of $Q\left(X_2^-\right)$ in neighboring cells.
Moreover, the most widely considered domain structure, whose $X_2^-$ mode switches sign at the domain wall, has a much larger domain wall motion energy compared with domain structures that have consistent $X_2^-$ mode.
This understanding based on lattice mode analysis deepens the broader understanding of the distinctive ferroelectric behavior of HfO$_2$.

\section*{acknowledgement}
This work was supported by ONR N00014-21-1-2107. 
First-principles calculations were performed using the computational resources provided by the Rutgers University Parallel Computing (RUPC) clusters and the High-Performance Computing Modernization Office of the Department of Defense.

\bibliography{cite}

\end{document}